\documentstyle[12pt,graphicx,amstex]{article}

\begin{document}

\begin{center} 
\begin{Large} 
{\bf Observable frequency shifts 
via  \\  spin-rotation coupling}
\end{Large}
\end{center}
\vspace*{3mm}
\begin{center} 
{\bf Bahram Mashhoon$^1$, Richard Neutze$^2$, 
Mark Hannam$^3$ } \\ 
{\bf and Geoffrey E. Stedman$^4$ } \\ 
\vspace*{8mm}
$^1$ {\it Department of Physics and Astronomy, University of 
Missouri-Columbia, \\ Columbia, Missouri 65211, USA}  \\
$^2$ {\it Department of Biochemistry, Uppsala University, Biomedical Centre, 
\\ Box 576,  S-75123 Uppsala, Sweden}  \\
$^3$ {\it Department of Physics and Astronomy, University of North Carolina,
Chapel Hill, NC 27599-3255, USA}\\
$^4$ {\it Department of Physics and Astronomy, University of Canterbury, \\
Private Bag 4800, Christchurch, New Zealand} \\
\end{center}
\vspace*{8mm}
\centerline{\large {\bf Abstract}} 
\vspace*{5mm}
The phase perturbation arising from spin-rotation coupling is developed as a 
natural extension of the celebrated Sagnac effect.   
Experimental evidence in support of this phase shift, 
however, has yet to be realized due to the exceptional sensitivity 
required.  We draw attention to the relevance of a series of experiments 
establishing that circularly polarized light, upon 
passing through a rotating half-wave plate, is changed in frequency 
by twice the rotation rate. These experiments may be interpreted 
as demonstrating the role of spin-rotation coupling in inducing
this frequency shift, thus providing direct empirical verification 
of the coupling of the photon helicity to rotation.  A neutron 
interferometry experiment is proposed which would be sensitive 
to an analogous frequency shift for fermions. In this arrangement, 
polarized neutrons enter an interferometer containing two 
spin flippers, one of which is rotating while the other is held stationary. 
An observable beating in the transmitted neutron 
beam intensity is predicted.

\newpage
Theoretical interest in the influence of rotation on the  
phase of light passing through an optical interferometer 
already dates over a century~\cite{lodge1893}. 
Sagnac's observation of a phase shift proportional to the scalar product 
of the rotation frequency 
and the area of his interferometer~\cite{sagnac1913} 
provided an empirical basis for a rich field of both fundamental and applied 
research into the influence of rotation on the phase of a quantum 
mechanical wave function~\cite{stedman1997}.

The Sagnac effect may be regarded as a manifestation of the 
coupling of orbital angular momentum of a 
particle, $\mbox{\boldmath$L$} = \mbox{\boldmath$r$} \times 
\mbox{\boldmath$p$}$, to rotation.   Suppose any radiation 
propagates in vacuum around a rotating interferometer and has frequency 
$\omega_0$ and wave vector $\mbox{\boldmath$k$}_0$
when measured in the corotating frame.  An inertial 
observer $O$ will observe that the
wave vector of the radiation along the 
$i$th arm of the interferometer is (at first order in $\Omega$)
$
\mbox{\boldmath$k$}_i = 
\mbox{\boldmath$k$}_0 + \omega_0 \; \mbox{\boldmath$\Omega$} \times 
\mbox{\boldmath$r$}_i/c^2  
$
such that a phase shift arises:
\begin{eqnarray}  \label{eqsagnac}
\Delta \Phi& =& \sum_i \Delta \mbox{\boldmath$k$}_i \cdot \Delta 
\mbox{\boldmath$r$}_i 
=\frac{ \omega_0}{c^2} \sum_i \mbox{\boldmath$\Omega$} \cdot 
\mbox{\boldmath$r$}_i  \times \Delta \mbox{\boldmath$r$}_i \nonumber \\ 
&=&  \frac{2 \omega_0 }{c^2} \; \mbox{\boldmath$\Omega$} \cdot 
\mbox{\boldmath$A$} 
=
\frac{1}{\hbar} \sum_i \mbox{\boldmath$\Omega$} \cdot 
\mbox{\boldmath$L$}_i \; \Delta t_i  \;\;\;, 
\end{eqnarray}
where we have used 
$\Delta \mbox{\boldmath$r$}_i = \Delta t_i \, 
\mbox{\boldmath$v$}_i$, for any particle in vacuum 
$\mbox{\boldmath$v$}_i = c^2 \mbox{\boldmath$k$}_i/\omega_i
= c^2 \mbox{\boldmath$p$}_i/\omega_i \hbar $, 
and $\mbox{\boldmath$A$} \equiv 1/2 \sum_i 
\mbox{\boldmath$r$}_i  \times \Delta \mbox{\boldmath$r$}_i $ is 
the area of the interferometer~\cite{dresdenET1979}.

The Sagnac phase shift \eqref{eqsagnac} is a scalar quantity that is
independent  
of the motion of the observer.  The same result therefore applies 
for an observer $O'$ at rest in the corotating frame. 
An interpretation of this expression for $O'$ is that the coupling of 
orbital angular momentum to rotation induces a frequency perturbation
(relative to that measured by $O$) 
proportional to $\mbox{\boldmath$\Omega$} \cdot \mbox{\boldmath$L$}$.  
Summing this frequency perturbation over the time of flight of a particle 
around the interferometer in effect recovers the Sagnac phase shift. 
From the standpoint of our rotating observer, Eq. (\ref{eqsagnac}) 
may naturally be extended to include the 
intrinsic spin of a quantum mechanical particle through replacing 
the orbital angular momentum $\mbox{\boldmath$L$}$  with the 
total angular momentum $\mbox{\boldmath $J$}=\mbox{\boldmath
$L$}+\mbox{\boldmath $S$}$.  This formalism consequently 
predicts that in the rotating frame, in addition to the Sagnac phase 
shift, a displacement of the interference fringes due to spin-rotation 
coupling will arise  proportional to 
\begin{equation} \label{eqSR}
\Delta \Phi_{\mbox{{\small SR}}} = 
\frac{1}{\hbar} \; \sum_i 
\mbox{\boldmath$\Omega$} \cdot 
\mbox{\boldmath$S$}_i  \:
\Delta t_i \;\;\; .
\end{equation} 
It has been shown how, with the addition of elements
which reverse the spin of a neutron~\cite{mashhoon1988}, 
or a photon~\cite{mashhoon1989}, along specific sections of an 
interferometer, a phase shift arises due to the coupling of spin to 
rotation which agrees with Eq.~(\ref{eqSR}). For a realistic experimental 
apparatus, however, such phase shifts are extremely small and this  has 
precluded their direct observation to date. 
In this letter, we draw attention to a series of closely related 
experiments which have provided 
empirical confirmation of helicity-rotation coupling for photons. 
Their experimental design allows a natural extension to neutron 
interferometry which we describe, enabling a direct interferometric 
test of spin-rotation coupling for fermions. 

The phenomenon of spin-rotation coupling is of basic 
interest since it reveals the inertial properties of intrinsic spin.  
In the formal realization of the invariance of quantum systems under 
inhomogeneous Lorentz transformations, the irreducible
unitary representations of the inhomogeneous Lorentz group are 
indispensable for the description of physical states.  These
representations are characterized by means of mass and spin.  
The inertial properties of mass in moving frames of reference 
are already well known: for instance via Coriolis, centrifugal 
and other mechanical effects~\cite{moorheadET1996}.  
The coupling of intrinsic spin
with rotation reveals the rotational inertia of intrinsic spin.

The underlying physics of spin-rotation coupling may intuitively be 
illustrated through a thought experiment.  Imagine our observer, 
$O'$, rotates with angular velocity $\mbox{\boldmath$\Omega$}$ parallel 
to the direction of  propagation 
of a plane linearly polarized monochromatic electromagnetic
wave whose electric field can be 
written in the coordinates of a reference inertial frame 
$I$ as: 
\begin{equation}  \label{eqlinear}
\mbox{\boldmath $E$} = E_{0} \; 
{\bf \hat{x}} \: e^{-i \omega t + i k z} ,
\end{equation} 
where $E_{0}$ is a constant amplitude, ${\bf k} = k {\bf \hat z}$ 
is the wave vector and $\omega= ck$. 
The coordinate system of $O'$ is related to  $I$ by
\begin{eqnarray} \label{eqcoord}
{\bf \hat{x}}+i {\bf \hat{y}}=e^{\pm i \Omega t} \left ( 
{\bf \hat{x}}^{\prime} \pm i {\bf \hat{y}}^{\prime}
\right ); \; \; \; \; {\bf \hat{z}}={\bf \hat{z}}'; \; \; \; \;  
t=t',
\end{eqnarray} 
such that,  from the viewpoint of the rotating observer,
\begin{equation}
\mbox{\boldmath $E$}=E_{0} \left ({\rm cos} \Omega t \: 
{\bf \hat{x}}^{\prime}-{\rm sin} \Omega t \: {\bf \hat{y}}^{\prime}
\right ) e^{-i \omega t + ikz},
\end{equation} 
with the direction of linear polarization appearing to 
rotate in a clockwise sense about the $z-$axis.  Our treatment ignores 
certain relativistic complications and focuses attention instead on the 
simple fact that from the viewpoint of the rotating observer, $O'$, 
the direction of linear polarization that is fixed in the inertial 
frame $I$ must drift in a clockwise sense with frequency $\Omega$ 
about the direction of propagation.   

Linearly polarized light represents a coherent superposition of 
right circularly polarized (RCP) and left circularly polarized (LCP) waves,
\begin{equation}
\mbox{\boldmath $E$}=\frac{1}{2}E_{0} \left ({\bf \hat{x}}+i{\bf \hat{y}} 
\right ) e^{-i \omega t+ikz}+\frac{1}{2}E_{0} \left
({\bf \hat{x}}-i{\bf \hat{y}} \right ) e^{-i \omega t+ikz}.
\end{equation}
From the viewpoint of the rotating observer the radiation field may also 
be written as a sum of RCP and LCP components
\begin{equation}
\mbox{\boldmath $E$} = \frac{1}{2} E_{0} \left ( {\bf \hat{x}}^{\prime}+i {\bf 
\hat{y}}^{\prime} \right ) e^{-i(\omega-\Omega)
t+ikz}+\frac{1}{2} E_{0} \left ({\bf \hat{x}}^{\prime}-i{\bf \hat{y}}^{\prime} 
\right ) e^{-i(\omega+\Omega)t+ikz},
\end{equation} 
as these eigenstates of the radiation field remain invariant 
under rotation but their frequencies undergo the characteristic `Zeeman' 
splitting that has a simple physical interpretation.  In a RCP (LCP) 
wave, the electric and magnetic fields rotate in the positive 
(negative) sense about the direction of propagation with frequency 
$\omega$.  Since the observer rotates in the positive sense with 
frequency $\Omega$, it perceives the effective frequency 
of the RCP (LCP) wave to be $\omega -\Omega$ ($\omega + \Omega$) 
with respect to time $t$.  
The proper time along the worldline of $O'$ is $\tau =
t/\gamma$, where $\gamma$ is the Lorentz factor  
such that the frequencies of the RCP and LCP light as measured by 
$O'$ are 
\begin{equation}  \label{eqomega}
\omega^{\prime}=\gamma(\omega \mp \Omega).
\end{equation} 
In this expression the Lorentz factor accounts for time dilation, 
which is consistent with the transverse Doppler effect.  In addition 
`angular Doppler terms' ($\mp \gamma \Omega$) arise due to the 
observer's rotation. Writing Eq.\ (\ref{eqomega}) in terms of energy 
as ${\cal E}^{\prime}=\gamma({\cal E}\mp \hbar \Omega)$ illustrates 
that the deviation from the simple transverse Doppler effect 
stems from the coupling of the spin of a circularly polarized photon 
to the rotation of the observer, since a RCP (LCP) photon carries an 
intrinsic spin of $\hbar (- \hbar)$ along its direction of 
propagation~\cite{beth1936}.  

Now replace the concept of a rotating observer that measures the 
frequency components of circularly polarized light with 
the atoms constituting a slowly rotating half-wave plate (HWP). 
Suppose RCP light falls perpendicular to the surface of   
this optical element, illustrated in Fig.\ 1.\  
Eq.\ (\ref{eqomega}) now describes the frequency 
of the radiation that drives the motion of electrons within 
this material.  As such, RCP light will cause electrons to oscillate 
with frequency $\omega'_{RCP} \approx \omega - \Omega$ 
in the rotating frame.  Furthermore, the action of the HWP is to 
transform RCP to LCP light such that light transmitted through the 
HWP will become LCP and will have the same frequency, 
$\omega_{LCP}' \approx \omega - \Omega$,  {\it in the 
rotating frame of reference}. Through the inverse transformation 
of Eq.\ (\ref{eqomega}), {\it i.e.} $\omega_{LCP} 
\approx \omega_{LCP}' - \Omega$, 
the transmitted light in $I$ will be both LCP {\it 
and shifted in frequency by} 
\begin{equation} \label{eqfreqshift}
\Delta \omega_{RCP \rightarrow LCP} = - 2\, \Omega, 
\end{equation}
hence the medium absorbs energy, linear momentum and angular
momentum from the radiation field.   Conversely, for LCP
radiation passing through the same system the relevant spin-rotation 
frequency shift would involve $\omega^{\prime}_{LCP} \approx 
\omega + \Omega$ such that 
$\Delta \omega _{LCP \rightarrow RCP} = 2\Omega$.  It follows that for 
linearly polarized light no net transfer of energy, momentum or angular 
momentum occurs!

These results can be extended to more general spin states via the
superposition principle. For instance, if in Fig.\ 1 the rotating HWP
is replaced by a rotating quarter-wave plate, the outgoing radiation
will be a superposition of a RCP component with frequency $\omega$ and
a LCP component with frequency $\omega-2\Omega$. 

Identical conclusions to these heuristic arguments 
have been drawn from Maxwell's equations when considered in the rotating 
frame of reference~\cite{mashhoon1989,pippard1994}. Furthermore, and of 
central importance to this letter, a series of experiments have been 
performed which confirm the frequency shift predicted above.  
A helicity-dependent rotational frequency shift was first 
observed using microwave 
radiation~\cite{allen1966}.  This effect has subsequently been 
investigated in the optical regime by several authors~\cite{garetzET1979}.  
These studies provide direct experimental verification of the phenomenon 
of helicity-rotation coupling for electromagnetic radiation.

With these experimental results in mind, the connection between the 
{\it frequency shift}, Eq. (\ref{eqfreqshift}), and the {\it 
constant optical phase shift} predicted 
by Eq.~(\ref{eqSR}), can be clarified in a simple configuration 
as follows.  Let an optical interferometer 
be set in rotation as illustrated in Fig. 2.  When viewed 
from $I$, RCP light having passed through the first HWP becomes LCP 
and is shifted in frequency by $-2 \Omega$, 
Eq.~(\ref{eqfreqshift}), as the HWP 
rotates with the interferometer at angular velocity 
$\mbox{\boldmath$\Omega$}$.  Multiplying this frequency shift by the 
time of flight of a photon between the two HWP's, $\Delta t = 
l/c$, gives in effect what amounts to a helicity-rotation phase 
shift $\Delta \Phi = \oint \mbox{\boldmath$k$}\cdot d \mbox{\boldmath$r$} 
= (\omega_+ - \omega_-) l/c = 2 \Omega \, l/c$, where $\omega_+ = \omega$ 
and $\omega_- = \omega -2\Omega$.   
This same phase shift is expected at the detector in the rotating 
frame and is given by Eq. (\ref{eqSR}), the factor of two arising as 
$ \mbox{\boldmath$\Omega$} \cdot \mbox{\boldmath$S$} = \pm \hbar $ 
for RCP or LCP light, respectively, in the rotating frame.   

We have considered thus far the simplest configuration for the
measurement of frequency shifts due to helicity-rotation
coupling, since the direction of propagation has been along the axis
of rotation.  The general expression for spin-rotation
coupling relating the energy  measured by a rotating observer to 
measurements performed in $I$ can be written as
\begin{equation} \label{eqgeneral}
{\cal E}^{\prime}=\gamma ({\cal E}-\hbar M \Omega),
\end{equation} 
where $M$ is the total (orbital plus spin) `magnetic' 
quantum number along the axis of rotation; that is,
$M=0, \pm 1, \pm 2, \ldots$ for a scalar or a vector field while 
$M \mp \frac{1}{2}=0, \pm1, \pm2, \ldots$ for a Dirac field. 
In the JWKB approximation, Eq. (\ref{eqgeneral}) can be written as
$ {\cal E}^{\prime}=\gamma({\cal E}- \mbox{\boldmath $\Omega$}
\cdot \mbox{\boldmath 
  $J$})=\gamma({\cal E}- \mbox{\boldmath $v$}\cdot \mbox{\boldmath
  $p$})-\gamma 
\mbox{\boldmath $S$}\cdot \mbox{\boldmath
$\Omega$}$, 
so that in the absence of intrinsic spin we recover the
classical expression for the energy of a particle as
measured in the rotating frame with 
$\mbox{\boldmath $v$}= \mbox{\boldmath $\Omega$} \times 
\mbox{\boldmath $r$}$.  Spin-rotation coupling, however,  
violates the underlying assumption of locality in special relativity:
that the results of any measurement performed by an accelerating 
observer (in this case the measurement of frequency) are 
locally equivalent to those of a momentarily comoving inertial observer,
but agrees with an extended form of the locality hypothesis.  This is 
a nontrivial axiom since there exist definite acceleration scales of 
time and length that are associated with an accelerated observer.  
Discussion of this extension to the standard Doppler formula, 
and its wider implications on the theory of relativity, 
have been presented 
elsewhere~\cite{mashhoon1988,mashhoon1989,locality,mashhoon1995}.

Observational support for this energy shift for fermions has been 
provided via a small frequency offset in high-precision  experiments 
due to the nuclear spin of Mercury coupling to the rotation 
of the Earth~\cite{mashhoon1995,winelandET1991}. 
More general experimental arrangements which test Eq.~(\ref{eqgeneral}) 
can also be envisioned.  In fact, an experimental 
configuration~\cite{nienhuis1996} recently demonstrated \cite{courtialET1998}  
that linearly polarized light, when prepared as an eigenstate 
of the orbital angular momentum operator 
$L_z$, also suffers a frequency shift upon passing through a rotating 
Dove prism.   These observations can be explained on the basis 
of Eq.~(\ref{eqgeneral}); moreover, it would be interesting to 
repeat such experiments using circularly polarized radiation 
in order to see the combined coupling of the orbital plus spin 
angular momentum of the field to rotation. 

In analogy with 
the observed frequency shift when circularly polarized light 
passes through a rotating HWP, Eq.~(\ref{eqgeneral}) indicates that 
similar experiments should be possible using polarized neutrons.
To this end let neutrons propagating through a uniformly rotating 
spin flipper be polarized with  their spin 
$| \kern-.16em \uparrow \, \rangle 
\parallel \mbox{\boldmath$\Omega$}$, 
as illustrated in Fig. 3.   
Repetition of the arguments leading to Eq. (\ref{eqfreqshift}) gives, 
in the JWKB approximation,  a frequency shift (measured in $I$) for 
the transmitted neutrons equal to  
\begin{equation}  \label{eqfreqneut}
\Delta \omega_{| \uparrow  \rangle 
\rightarrow | \downarrow  \rangle }
= -2 M \Omega = - \Omega,
\end{equation}  
as the incident neutron in state 
$| \kern-.16em \uparrow \, \rangle $ has $M=1/2$.  It is assumed here
that the average energy of the neutron in the spin flipper remains
constant, i.\ e.\ there is no intrinsic frequency shift associated
with the spin flipper; otherwise, the additional shift should also be
taken into account. Moreover, our result for the neutron frequency
shift can be extended to more general spin states. The simplest spin 
flipper consistent with our assumption would be a coil producing a
uniform static magnetic field  
$B$ normal to the polarization axis of the neutrons~\cite{allmanET1997}.  
If $t$ is the interval of time that it takes neutrons of speed $v_{n}$ 
to traverse the length of the coil, the probability of spin flip upon 
passage is $\rm{sin}^{2}\zeta$, where $\zeta=-\mu_{n}Bt/\hbar$ and 
$\mu_{n}$ is the neutron magnetic moment.  The neutron spin would 
therefore flip for $\zeta=(2N+1)\pi/2$, $N=0,1,2,\ldots$.  The length 
$L$ of an appropriate coil can thus be obtained from $L=(2N+1)L_{0}$, 
where $L_{0}=v_{n}t$ for $\zeta=\pi /2$; hence, 
$L_{0}=\pi \hbar v_{n}/(2|\mu_{n}|B)$.  For $B=500 \; {\rm G}$ and 
$v_{n}/c \simeq
10^{-5}$, we find $L_{0}\simeq$ 1 mm; in this case, thermal neutrons of 
wavelength $\lambda \simeq 1$ \AA $\;$ would take less
than a $\mu$sec to traverse the coil.  Should the coil be rotated 
slowly, the various approximations involved in our treatment could be 
justified.

One can imagine a variety of interferometric configurations using 
rotating spin flippers. If the arrangement is such as to produce 
a constant phase shift then, in effect, such experiments would 
be similar to the configuration suggested a decade ago~\cite{mashhoon1988}. 
Because this phase shift is very small a large-scale neutron 
interferometer is required for its possible realization.  
It is therefore interesting to conceive an interference
experiment that would observe a beating between two different 
de Broglie frequencies.  A beat frequency of 
$\simeq 2 \times 10^{-2}$ Hz has previously  been measured in a 
neutron interferometry experiment~\cite{badurekET1986} involving the 
passage of neutrons through stationary rf coils driven at slightly
different frequencies.

As illustrated in  Fig.\ 4, we propose to place identical 
spin flippers along each of the two separated neutron beams such 
that an intensity maximum is recorded when both coils are aligned 
parallel.  Keeping the interferometer stationary in the inertial frame 
of the laboratory, $I$, we then rotate one of the coils with angular
velocity $\mbox{\boldmath$\Omega$}$ parallel to the neutron wave
vector.  From Eq. (\ref{eqfreqneut}), a shift in  
frequency of this beam by
$\Delta \omega = - \Omega$ will be induced such that a time-dependent 
interference intensity envelope of the form $I \propto \left [ 1+
\cos \left (\Omega t + \phi_0 \right  ) \right ]$ arises, 
where $\phi_{0}$ is the  constant phase shift between the two 
interferometer arms~\cite{phase}.   The frequency components of this 
intensity modulation may easily be recovered by recording the 
intensity as a function of time and taking the Fourier transform of 
the output~\cite{neutzeET1998}.  A sinusoidal modulation of the intensity 
arising from spin-rotation coupling will cause sideband structure to 
appear in the resulting spectra, with the peaks separated by $\Omega$, 
the rate of rotation of the spin-flipper.  
As the proposed experimental apparatus closely resembles that used 
by Allman {\it et al.}~\cite{allmanET1997} 
to measure simultaneously geometric and dynamical phase shifts, 
we believe  that the potential  
observation of a spin-rotation coupling induced frequency shift 
for fermions falls entirely within the sphere of current technology.

All of the experimental work to date has involved rotation frequencies 
$\Omega \ll \omega$ and the interpretation of the experiments has been 
based on certain intuitive 
considerations~\cite{pippard1994,allen1966,garetzET1979,nienhuis1996}.  
The present identification of the origin of these results 
in terms of spin-rotation coupling makes it possible to discuss the 
general situation for arbitrary $\Omega$  and spin, as well as whether RCP 
radiation can stand completely still for $\omega = \Omega$ in Eq.\
\eqref{eqomega}. 
This situation is reminiscent of the pre-relativistic Doppler formula 
for linear motion, which predicted that electromagnetic radiation would 
stand completely still relative to an observer moving with 
speed $c$ along the direction of propagation of the wave.  This
circumstance proved an important   
motivating factor in Einstein's development of the theory of 
relativity~\cite{einstein}.
These issues have been the subject of a number of theoretical 
investigations and it is hoped that 
further experimental studies can shed light on future developments 
toward a nonlocal theory of accelerated systems~\cite{mashhoon1993}.

\subsection*{Acknowledgements}
We wish to thank A.I. Ioffe, H. Kaiser, and S.A. Werner for
discussions regarding the neutron interferometry experiments suggested 
in this letter and Jenny C.\ Williams for discussions on spin-rotation
coupling.

\newpage
\centerline{\large\bf Figure Captions} 
\renewcommand{\labelenumi}{Fig.\ \theenumi.}
\begin{enumerate}
\item Frequency redshift via a rotating half-wave plate (HWP). 

\item Schematic plot of a rotating interferometer. A half-wave plate
flips the initial helicity along one path while a second half-wave
plate flips it back before recombination in order that interference
can take place. The distance between the HWPs is $l$. 

\item Schematic depiction of the passage of longitudinally polarized
neutrons through a uniformly rotating spin flipper. We assume that the
average energy of the neutron does not change while in the spin
flipper. The rotational energy shift, ${\cal E}_f - {\cal E}_i =
- \hbar \Omega$, provides a new way to moderate neutrons. Note that if
the sense of rotation is reversed, then there would be a gain in
energy by $\hbar \Omega$. 

\item A neutron interferometer in an inertial frame of
reference. Longitudinally 
polarized neutrons pass through a slowly rotating spin flipper along
one arm and a static spin flipper along the other arm resulting in a
beat phenomenon at the detector.
\end{enumerate}

\includegraphics{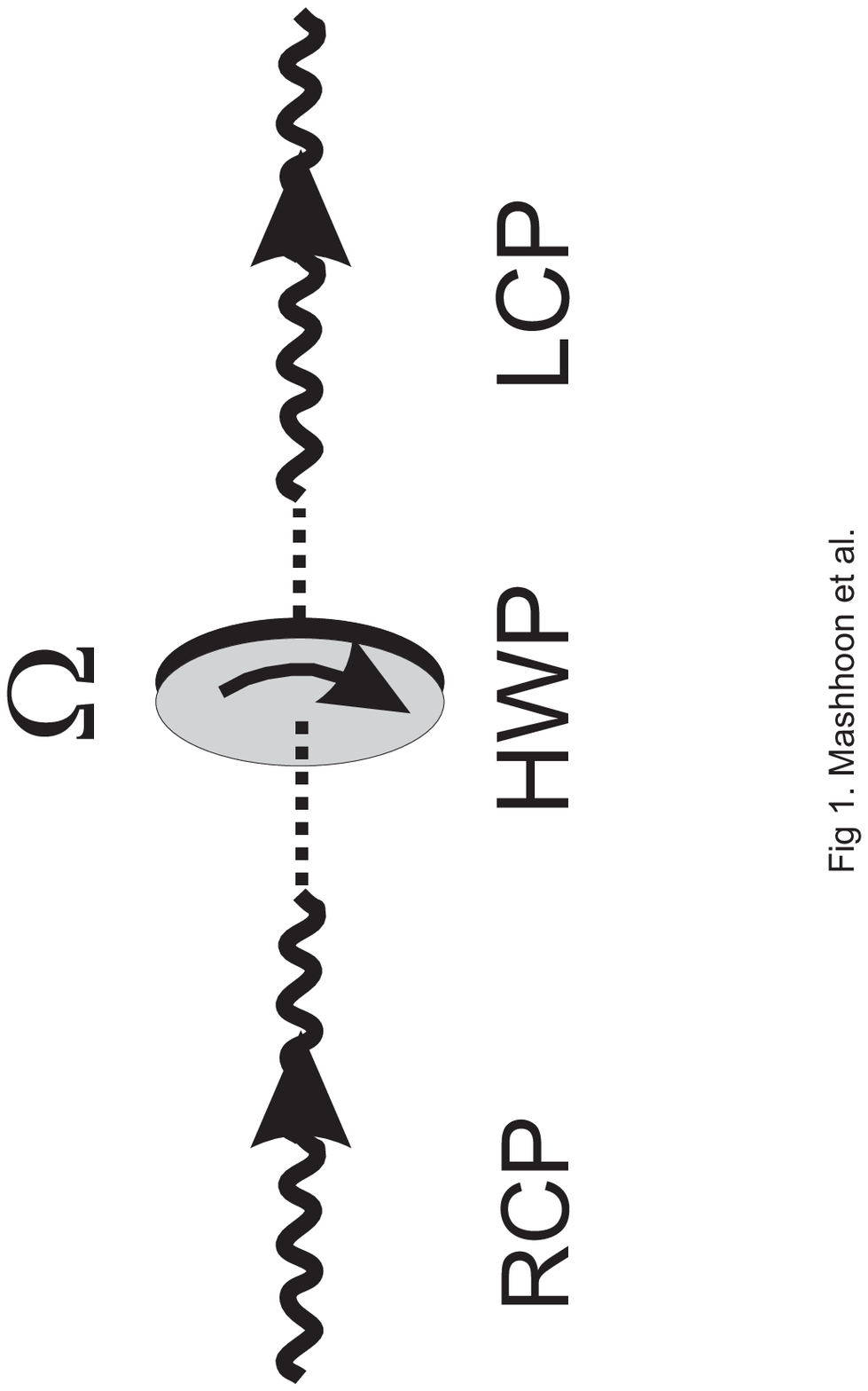}

\includegraphics{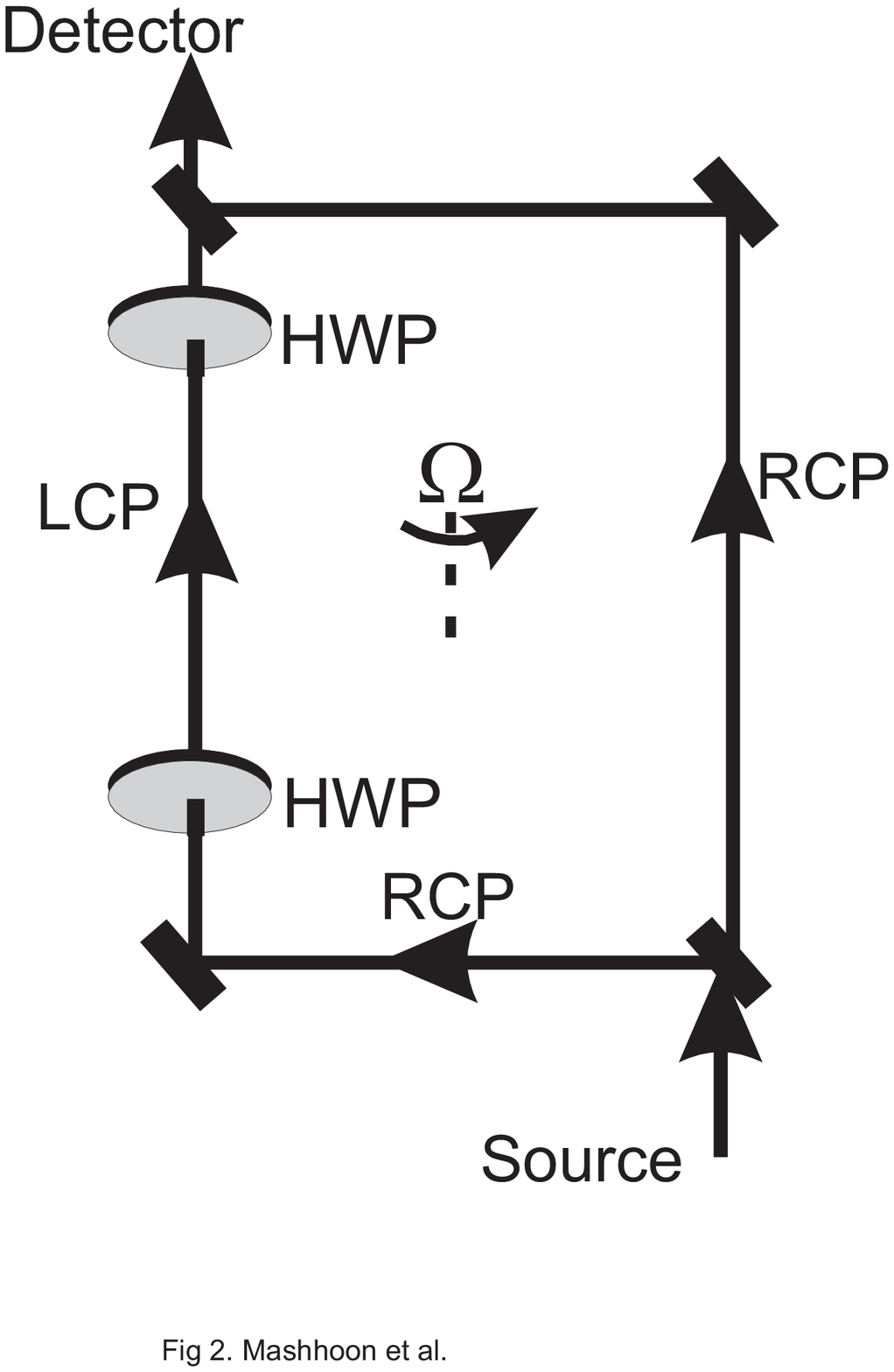}

\includegraphics{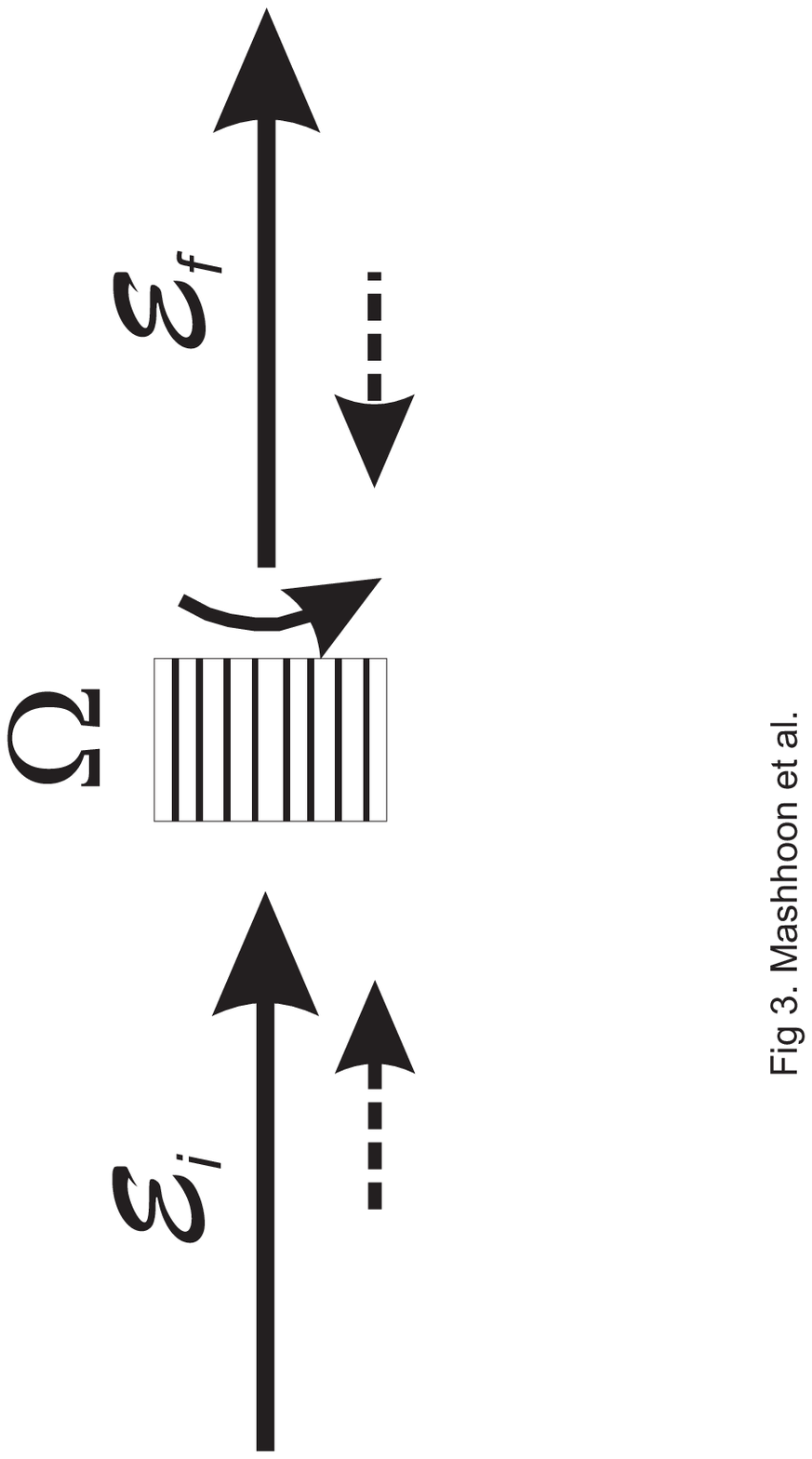}

\includegraphics{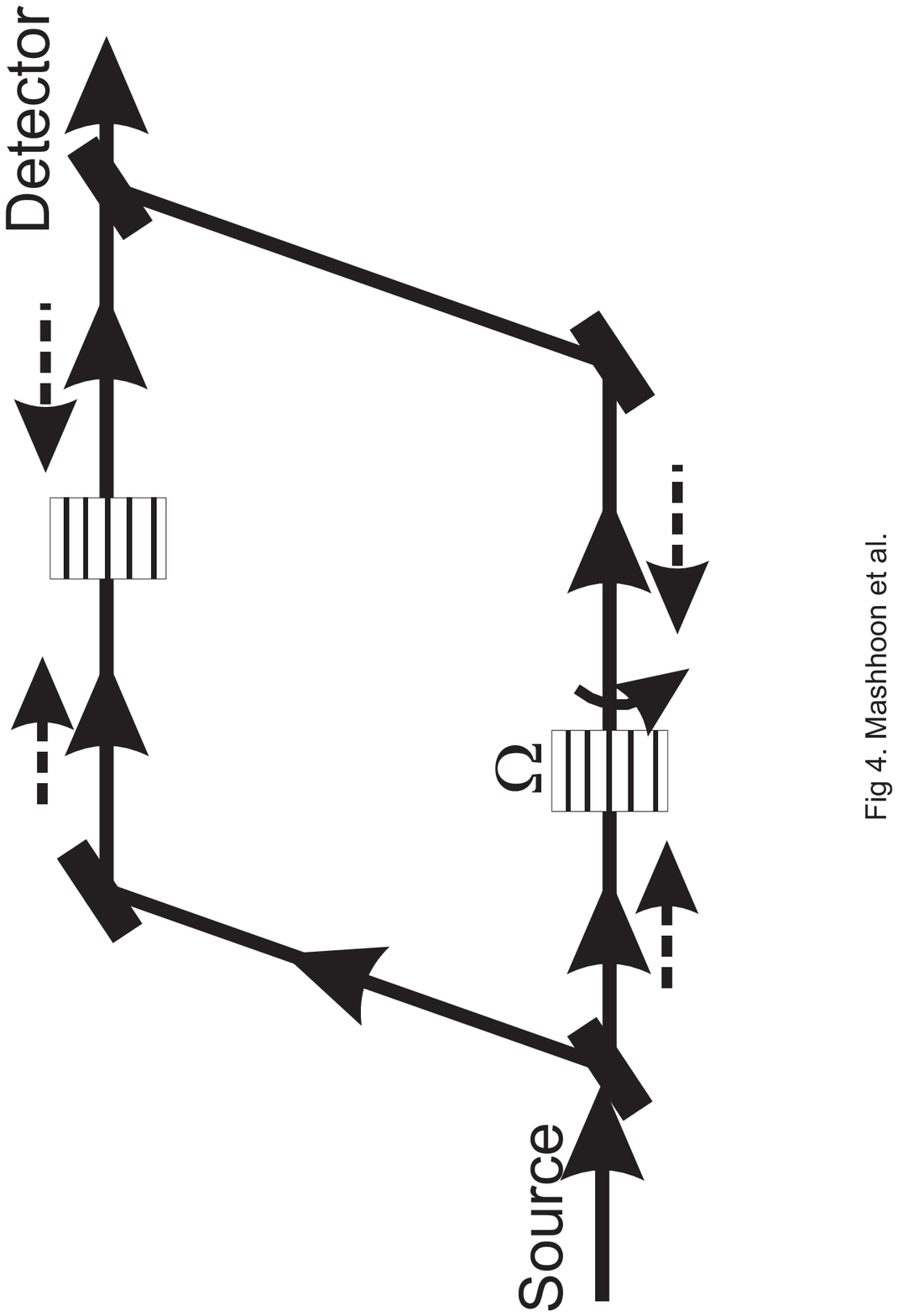}


\begin{thebibliography}{999}
\bibitem{phase}
The phase shift $\phi_{0}$ depends on $\Omega$ as well; in fact, in 
the absence of other contributions $\phi_{0}=-\Omega d / v_{n}$, 
where $d$ is the interferometric distance between 
the rotating spin flipper and the detector.
\end{thebibliography}
\end{document}